\documentclass[12pt]{article}

\newenvironment{wileykeywords}{\textsf{Keywords:}\hspace{\stretch{1}}}{\hspace{\stretch{1}}\rule{1ex}{1ex}}

\usepackage{amsmath,amssymb}
\usepackage{graphicx}
\usepackage{color}
\usepackage{dcolumn}
\usepackage{bm}
\usepackage[numbers,super,comma,sort&compress]{natbib}

\definecolor{background-color}{gray}{0.98}

\title{Implementing Dimer Metadynamics using GROMACS}
\author{M Nava\thanks{Department of Chemistry and Applied Biosciences, ETH Zurich, and Facolt\`a di Informatica, Istituto di Scienze Computazionali, Universit\`a della Svizzera Italiana, Via G. Buffi 13, 6900 Lugano
 Switzerland}}

\newcommand{\onlinecite}[1]{\hspace{-1 ex} \nocite{#1}\citenum{#1}} 

\begin{document}

\maketitle

\begin{abstract}
We develop a Gromacs implementation of Dimer Metadynamics[JCTC 13, 425 (2017)] for enhanced sampling through artificial delocalization effects. 
This implementation is based entirely on a Plumed collective variable developed for this purpose, 
the fine tuning of Gromacs input parameters, modified forcefields and custom non-bonded interactions. We demonstrate 
this implementation on Alanine Dipeptide in vacuum and in water, and on the 12-residue 
Alanine Polypeptide in water and compare the results with a standard multiple-replica technique such as Parallel Tempering. 
In all the considered cases this comparison is consistent and the results with Dimer Metadynamics are smoother and 
require shorter simulations, thus proving the consistency and effectiveness of this Gromacs implementation.

\end{abstract}

\begin{wileykeywords}
GROMACS, Replica Exchange, Enhanced Sampling,Protein Simulation,Dimer Metadynamics
\end{wileykeywords}

\clearpage


\begin{figure}[h]
\centering
\fbox{
\begin{minipage}{1.0\textwidth}
\includegraphics[width=110mm,height=20mm]{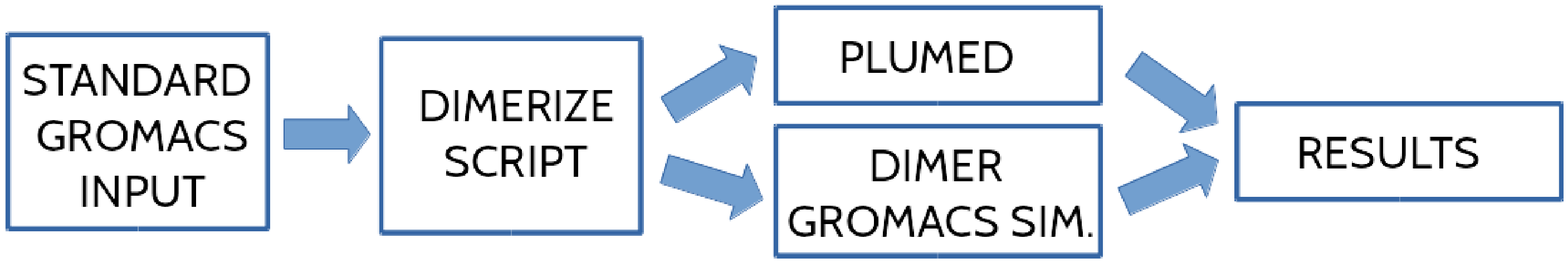}
\\
Gromacs is a popular program for the simulation of proteins using Molecular Dynamics techniques. In this work 
we present a Python script that enables Gromacs to simulate proteins with Dimer Metadynamics, an enhanced sampling method that 
can be used when a system has different important configurations that are difficult to explore with standard methods because large potential energy barriers 
have to be overcome. We demonstrate the implementation with three examples of application on proteins that are commonly used as benchmark for 
new methodologies.
\end{minipage}
}
\end{figure}

  \makeatletter
  \renewcommand\@biblabel[1]{#1.}
  \makeatother

\bibliographystyle{apsrev}

\renewcommand{\baselinestretch}{1.5}
\normalsize

\clearpage

\section*{\sffamily \Large INTRODUCTION} 

It is a fact that Nature has different timescales\cite{timescales}. This is well known to Gromacs\cite{grom} users who often deal with systems characterized by free energy surfaces 
with several metastable states separated by high potential barriers. A typical example is the simulation of proteins, where conformational changes 
can happen on timescales several orders of magnitude longer than the thermal motion of the atoms. Sampling methods based on brute force alone are 
unpractical on most of these systems\cite{longscale1,longscale2,longscale4,meta_review}. 

To overcome this problem many enhanced sampling methods have been developed over the years\cite{met1,met2,meta_pnas,wtm,omar2}. 
Some of these methods work by enhancing the fluctuations of one or more arbitrarily chosen Collective Variable (CV) that are known to well represent the conformational changes of the system; other methods 
instead focus on a more general approach that has no need of {\it a-priori} knowledge of the system but relies on a series of parallel simulations, each with different Hamiltonians; these 
Replica Exchange methods\cite{re1,re2,re3,wte,re4,bussi_cvt,re6,re7,re8,re9,re10} are a generalization of Parallel Tempering\cite{ptempering} (PT), a method where each replica runs at a different temperature. 
Replicas at higher temperature have access to a 
greater volume in phase space and are more likely to sample configurations that are normally hard to obtain at the target temperature, these configurations are then propagated to the replica at the 
target temperature with Monte Carlo moves. In this category there are also methods based on the enhancement of sampling through quantum effects\cite{Voth_sampling,quantumruge,dbms}. 
Here the enhancement of sampling is given by zero point energy 
that reduces the effective height of the barriers and by tunneling that can connect different metastable states with new, classically forbidden, pathways. 

One method in particular\cite{dbms} makes use of replicas that 
are at the same temperature but at different de Broglie wavelengths, so that the magnitude of quantum effects can be controlled and used as a sampling mechanism. 
In order to control the de Broglie wavelength, each replica is a quantum simulation based on Feynman's path integrals and indeed the method is suitable for both quantum and classical systems. Recent work\cite{gpi} 
focuses more on classical systems and builds from the idea that delocalization effects don't necessarily have to be described by Quantum Mechanics. Such effects can instead be introduced artificially 
with a classical simulation of dimers, whose binding interaction can be modified in different replicas to obtain dimers that can be either strongly or weakly bound. Loose dimers describe a system with 
strong pseudo-quantum effects and are used for sampling; these configurations are passed down to replicas with more and more tightly bound dimers until the target replica can be made rigorously classical in a similar 
way as done in Ref. \onlinecite{Voth_sampling}.

This method, named Dimer Metadynamics (DM) has been shown to be effective on proteins in vacuum. In a recent work\cite{gpi} we claimed that DM is insensitive to the presence of an explicit solvent basing our assumption on the fact that  
in DM one can choose which atoms are to be represented as dimers and which are to be left as ordinary atoms and the Monte Carlo moves between replicas have an acceptation rate that depends only on the 
atoms that have been ``dimerized''. This means of course that a protein can be dimerized while leaving the solvent unmodified. In this work we demonstrate the feasibility of this approach; to do that we have 
developed a Python interface that allows Gromacs to carry out DM simulations with the help of the Plumed\cite{plumed} plugin. This interface is presented in detail in the following sections.

\section*{\sffamily \Large METHODOLOGY}
\begin{figure}
\includegraphics[width=0.8\columnwidth,keepaspectratio=true]{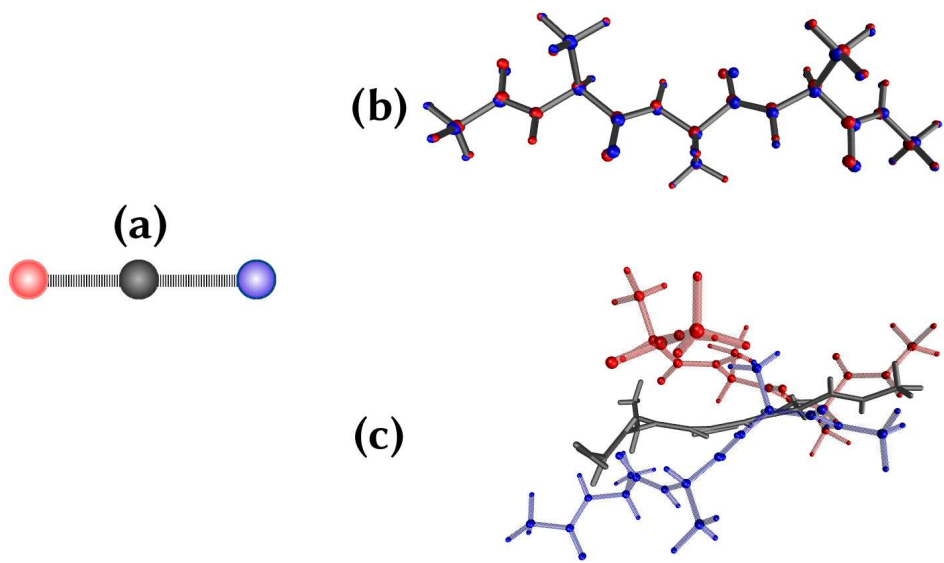}
\caption{\label{fig_dimers} (Color online) (a) The structure of a dimer. Blue and Red balls are the two beads composing the dimer, the gray ball in the middle is the position of 
the center of mass that in the Boltzmann replica represents the position of the physical particle. (b) A dimerized configuration of Alanine Dipeptide for small interaction 
parameter $\sigma$. The beads are localized and the typical shape of Alanine Dipeptide can be recognized from the position of the centers of mass. (c) The behavior at large $\sigma$, where 
the center of mass has no longer physical meaning and the two sets of beads (red and blue) are sampling the potential energy surface of Alanine Dipeptide at rescaled interaction.}
\end{figure}

The purpose of Dimer Metadynamics is to sample a Boltzmann distribution whose partition function is 
\begin{eqnarray}
Z_0 = \int dR \: \mbox{e}^{-\beta V(R)} \label{z01}
\end{eqnarray}
where $V(R)$ is the interaction potential, $\beta = 1/kT$ the inverse temperature and $R=\vec{r}_i, i = 1...N$ are 
the coordinates of the $N$ particles that make up the system. If we now introduce a new many-body coordinate $X = \vec{x}_i, i=1...N$ and  
functions $F_\sigma (X)$ that satisfy the relation $\int dX \: \mbox{e}^{-\beta F_\sigma (X)} = 1$, Eq. \eqref{z01} can be rewritten as 
\begin{eqnarray}
Z_0 = \int dRdX \: \mbox{e}^{-\beta V(R)}\mbox{e}^{-\beta F_\sigma (X)}
\end{eqnarray}
that with the transformation to coordinates $R_1,R_2$ $X=R_1-R_2$ and $R=(R_1+R_2)/2$ becomes
\begin{eqnarray}
Z_0 = \int dR_1 dR_2 \mbox{e}^{-\beta V\left(\frac{R_1+R_2}{2}\right)}\mbox{e}^{-\beta F_\sigma \left(R_1 - R_2\right)}   \label{H0}
\end{eqnarray}
If the functions $F_\sigma$ are so that their $\sigma \rightarrow 0$ limit is a Dirac's delta,
\begin{eqnarray}
\lim_{\sigma \rightarrow 0} \mbox{e}^{-\beta F_\sigma (R_1 - R_2)} = \delta(R_1-R_2)
\end{eqnarray}
$Z_0$ can be approximated as 
\begin{eqnarray}
Z_\sigma = \int dR_1 dR_2 \: \mbox{e}^{-\frac{\beta}{2}V(R_1)} \mbox{e}^{-\beta F_\sigma (R_1-R_2)} \mbox{e}^{-\frac{\beta}{2}V(R_2)} \label{Hsig}
\end{eqnarray}
where for $\sigma$ small enough $V(\frac{R_1+R_2}{2}) \simeq \frac{V(R_1)}{2} + \frac{V(R_2)}{2}$ and of course $\lim_{\sigma \rightarrow 0} Z_\sigma = Z_0$. 
This partition function is that of a system of dimers with a binding interaction $F_\sigma$ and interacting with each other with a rescaled potential $V(R)/2$. In Figure \ref{fig_dimers} 
we show the behavior of the system of dimers in the case of Alanine Dipeptide. For 
large $\sigma$ the dimers are delocalized and the two realizations of the system, respectively with coordinates $R_1$ and $R_2$ are free to explore rather different 
configurations. On the other hand, for small $\sigma$ the dimers are tightly bound and the Boltzmann behavior is recovered. 

The binding interaction $F_\sigma$ is taken in the form 
\begin{eqnarray}
F_\sigma(R_1,R_2) = \sum_{i=1}^{N} f_\sigma \left(\vec{r}_i^{\,1}-\vec{r}_i^{\,2}\right) \label{fdec}
\end{eqnarray}
where $\vec{r}_i^{\,1}$ and $\vec{r}_i^{\,2}$ are the coordinates of the $i-$th atom that has been represented as a dimer composed of two beads 1 and 2 as described in Eq. \eqref{Hsig}. 
As a choice of $f_\sigma$ in this Dimer implementation we use 
\begin{eqnarray}
f_\sigma^{q}(r) = \left(1 + \frac{r^2}{2q\sigma^2}\right)^q - 1 \label{fsig}
\end{eqnarray}
where $q$ is a parameter that determines the asymptotic behavior of $f_\sigma^q$. For  $q=1$ the standard spring energy interaction of the Path Integral formalism\cite{feynmanhibbs} is recovered, for 
$0 < q < 1$ the behavior of the function is quadratic for small $r$ and grows slower at larger distances, for $q=1/2$ the long-$r$ behavior is linear. The $\sigma$ parameter controls the 
transition between the low-$r$ and large-$r$ regimes. 

Dimer Metadynamics is based on Replica Exchange. The replicas have partition functions $Z_i$ of the kind of Eq. \eqref{Hsig}, $Z_0$ represents the Boltzmann replica with $\sigma=\sigma_0$, 
the following replicas have progressively increasing values of $\sigma$ that we denote as $\sigma_0=\sigma_1< \sigma_2 ... < \sigma_M$. Each system is run in parallel and 
periodically a Metropolis move attempts to swap the configurations between neighboring replicas\cite{gpi}. The probability to accept these swaps is 
\begin{eqnarray}
p_{i,i+1} = \min \left[1, \mbox{e}^{-\beta\Delta E} \right]
\end{eqnarray}

where:
\begin{eqnarray}
\Delta E = \left[F_{\sigma_i}(R_1^{i+1} - R_2^{i+1})+F_{\sigma_{i+1}}(R_1^{i} - R_2^{i})\right] - \left[F_{\sigma_i}(R_1^{i} - R_2^{i})+F_{\sigma_{i+1}}(R_1^{i+1} - R_2^{i+1})\right] \label{deltaE}
\end{eqnarray}

note that the superscript in the manybody coordinates here denotes the replica index. 
In the special case of a swap between the Boltzmann replica and the $\sigma_1$ replica the probability of acceptation is:

\begin{flalign}
p_{0,1}=\min\left[1,\mbox{e}^{-\beta\Delta V^{(0,1)}}\right] \label{pacccg}
\end{flalign}
with:
\begin{flalign}
\Delta V^{(0,1)} = \left[ \frac{V(R_1^0)+V(R_2^0)}{2} + V\left(\frac{R_1^1+R_2^1}{2}\right)\right] - \left[ \frac{V(R_1^1)+V(R_2^1)}{2} + V\left(\frac{R_1^0+R_2^0}{2}\right)\right]
\end{flalign}
 
where we remark again that $\sigma_0 = \sigma_1$. The Boltzmann replica is described by the partition function of Eq. \eqref{H0} 
and thus its $\sigma$ value is irrelevant to the sampling.
As a last step, this scheme is enhanced with Well-Tempered Metadynamics\cite{wtm}, where the dimer binding energy per particle is used as CV:

\begin{eqnarray}\label{scv}
s=\frac{1}{N \beta}\sum_{i=1}^{N}\left[\left(1+\frac{\left(\vec{r}_i^{\,2}-\vec{r}_i^{\,1}\right)^2}{2q\sigma^2}\right)^q-1\right]
\end{eqnarray}

The effect of Metadynamics is to enhance the fluctuations of $s$, so that configurations where the dimer is more delocalized are more likely to occur. This in turn increases the probability 
to accept the swap between replicas and practically reduces the number of replicas required for a Dimer simulation. If $p(s)$ is the probability distribution of $s$, the effect of Metadynamics is 
in fact to sample a biased distribution $p_b(s) = \left[p(s)\right]^{1/\gamma}$ where $\gamma$ is a boosting factor\cite{wte} chosen as input. This increases the overlap between the $p(s)$ of neighboring replicas 
giving smaller values of $\Delta E$ in Eq. \eqref{deltaE}.

\section*{\sffamily \Large Implementation details}

The implementation of Dimer Metadynamics on Gromacs consists of two parts. The Plumed plugin introduces the Dimer interaction $F_\sigma$ with a 
collective variable and the Dimerizer script prepares a suitable Gromacs input, taking care of the Dimer structure and opportunely modifying 
the provided forcefield. The dimerizer package also comes with the ``tune\_replicas'' utility that helps the user in the choice of the values of $\sigma$ 
for the non-classical replicas. The proposed values are to be considered as a rough estimate to be fine tuned by hand and are estimated 
from the distribution of the dimer binding energy of a non-interacting system of dimers, taking also in consideration the Metadynamics boosting factor $\gamma$.

\begin{figure}
\includegraphics[width=0.8\columnwidth,keepaspectratio=true]{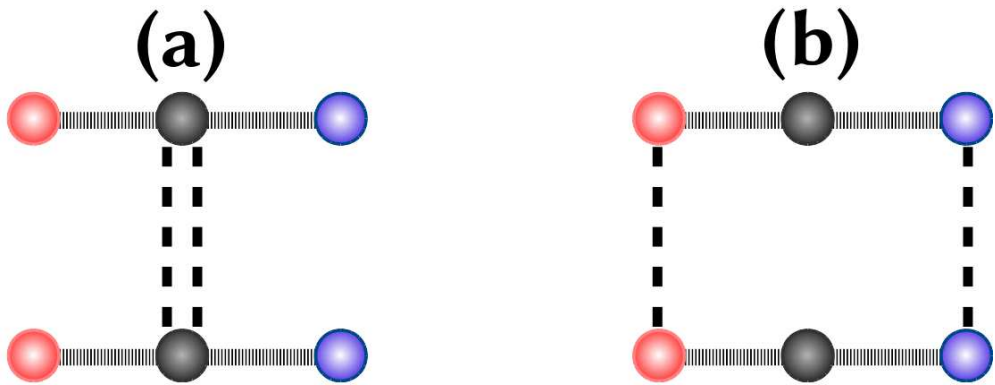}
\caption{\label{fig_inter} The structure of the interactions between dimers. (a) For the Boltzmann replica the beads interact via the full interaction potential $V(R)$ evaluated 
at the center of mass of the Dimer. (b) For the other replicas each bead interacts with a halved interaction potential $V(R)/2$. In all cases beads of different indices have only the binding 
interaction of the Dimer but otherwise do not interact with each other.}
\end{figure}

In Figure \ref{fig_inter} we review the shape of the interactions between dimers. Each Dimer consists of two beads and has a center of mass that, following the Gromacs notation, 
will be referred here as `virtual site', for reasons that will soon be clear. The only interaction that operates between beads of different indices (marked red and blue in the figure) 
is the Dimer binding interaction between beads belonging to the same Dimer. In any other case, red and blue beads do not interact directly. 
The Boltzmann replica, described by Eq. \eqref{H0} has the interaction potential evaluated on the virtual sites of each dimer (see Figure \ref{fig_inter}a), while in the other replicas 
the interaction is shared with equal weights on red and blue beads, as illustrated in Figure \ref{fig_inter}b. 

\subsubsection*{\sffamily \normalsize Plumed-CV}
The Dimer interaction has been implemented in the shape defined by Eq. \eqref{fdec} and Eq. \eqref{fsig}. Among the other possibilities, we have chosen to implement this interaction as a 
linear restraint in the Plumed plugin. A restraint in Plumed is in fact defined as a potential energy term of the shape $k(x - a)^2 /2 + m(x - a)$, where the argument $x$ here denotes a collective variable. 
The dimer interaction can thus be easily introduced with the ``RESTRAINT" command on the CV $s$ of Eq. \eqref{scv}, with $k=0$, $a=0$ and $m=1$:

\begin{verbatim}
RESTRAINT ARG=dim AT=0 KAPPA=0 SLOPE=1 LABEL=dimforces
\end{verbatim}

Where $dim$ is a reference to the Dimer CV that we have implemented in Plumed and is 
available in the stable branch of the official repository with the keyword ``DIMER". This is also the CV that has to be biased in order to enhance the Replica Exchange scheme. 
The format of the keyword is quite intuitive:

\begin{verbatim}
dim: DIMER Q=0.50 TEMP=300.00 DSIGMA=0.004 ATOMS1=d1b1 ATOMS2=d1b2
\end{verbatim}

where $Q$ is the $q$ exponent in Eq. \ref{scv} and $DSIGMA$ is the $\sigma$ value of the replica referred by the Plumed input file. Following the standard naming of Plumed input file, 
for each replica $n=0,...,M-1$ we use a $plumed.n.dat$ input file with specific settings. In particular each file with $n>0$ will have a different value of $\sigma$ that should be 
tuned in test-runs so as to optimize the swap rates. In order to avoid editing manually each of the Plumed input file we provide in the Dimerizer package also the `plread' script, a 
shell utility that updates all of the plumed input files with a single command. 
The Dimer CV also requires the atom indices (defined from $1$ to $N$) of the blue and the read beads, these are passed with the $ATOMS1$ and $ATOMS2$ arguments following the standard 
Plumed rules of enumerating atom groups.

\subsubsection*{\sffamily \normalsize Gromacs interface}
\begin{figure}
\includegraphics[width=0.8\columnwidth,keepaspectratio=true]{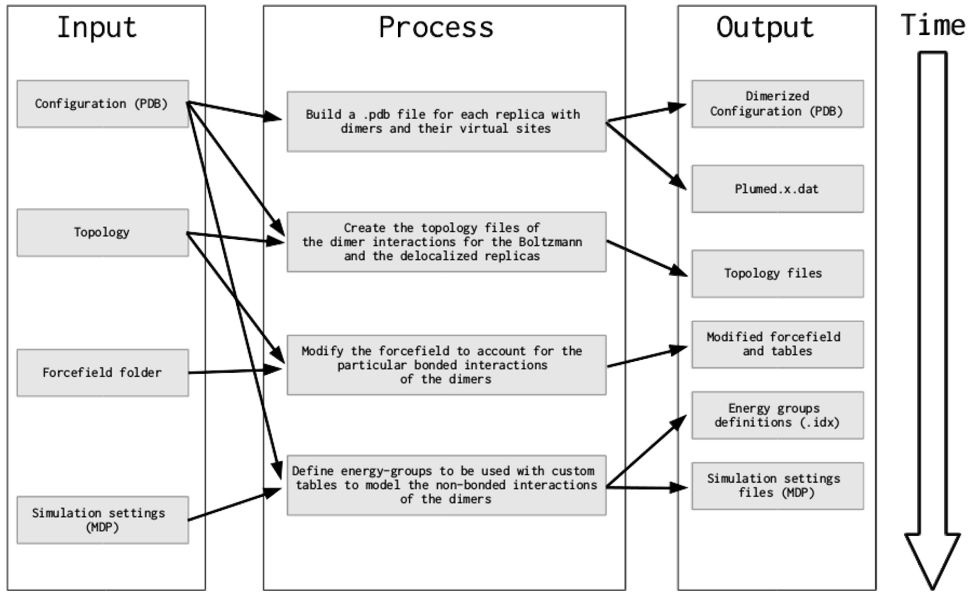}
\caption{\label{fig_flow} Flowchart representing the different stages of operations carried out by the dimerizer script.  Each 
stage uses one or more input files of a standard Gromacs simulation and produces output files for a Gromacs replica exchange simulation of dimers.
}
\end{figure}

The Plumed CV by itself is not enough to run a Dimer simulation in Gromacs as there are unconventional interactions that are not supported by the popular simulation tool. 
This difficulty can be overcome with customized forcefields and non-bonded interactions that are often not straightforward to implement. For this reason we have developed the 
Dimerizer script, whose function is that to take a standard Gromacs input and forcefield, and produce a Dimerized output with basic Plumed input files suitable for 
a Gromacs simulation. Given such an input there are several steps that are taken by the script and are shown in Figure \ref{fig_flow}. Firstly, the configuration file in Protein DataBank (PDB) format is converted 
to a PDB file of Dimers, where from the original file the missing beads and virtual sites are added for the atoms that have to be dimerized. For example, a $N=4$ atoms file where only 
atoms 1 and 2 are being dimerized becomes a $N=8$ file, where the Dimer beads have indices $d_1=(1,5)$ for the first dimer and $d_2=(2,6)$ for the second and the last two entries 
are the virtual sites. In this way only the lines of the atoms to be dimerized are duplicated  
while everything else, such as the solvent, is left unmodified. 

The topology file is then modified to represent interactions as in Figure \ref{fig_inter}: two topology files are thus produced as output, one for the Boltzmann replica (Figure \ref{fig_inter}a) and 
one for the other replicas (Figure \ref{fig_inter}b). The indices in the topology's interaction tables are redefined so as to point to either a bead or a virtual site 
for the dimerized atoms; for non-dimerized atoms, if the interaction line involves only other non-dimerized atoms it remains unchanged. 
The definition of the virtual sites is also added in the topology files in the section `[virtual\_sites2 ]'. 
Moreover, new atomtypes have to be defined in the `[ atoms ]' section in order to distinguish the 
interactions of the beads from those of the virtual sites and those of the non-dimerized atoms. This is necessary because the interactions with beads have to be halved and thus require the correct entry in the 
forcefield files. This is taken care of in the following stage, where from a given forcefield folder, modified forcefield files are given as output. This is done as follows: the script scans through the 
forcefield files looking for the entries pointed by the original topology file. When such an entry is found there are two possibilities: if the entry is never referred by dimerized atoms no action is taken, otherwise 
for each dimerized atomtype make the corresponding entry for the virtual site and for the beads, where in the latter the energy coefficients are halved. If in the forcefield some entries have wildcards, these are 
removed and the entries specific to the system are added instead. This step is required because putting into a forcefield entries for each possible combination of interactions is unpractical and produces 
forcefield files of about 2Gb of size that are unbearably slow to process by Gromacs. 

One last step is the manipulation of the non-bonded interactions, such as the Coulomb and the Lennard-Jones terms of the forcefield. 
Gromacs can compute these interactions independently in different energy groups that can be specified, moreover such interactions can be customized passing a user-defined interaction table between each of those groups. 
For a Dimer simulation, non-bonded interactions between beads of different color are null, interactions involving beads are halved and interactions involving virtual sites are full. 
This implies the use of three energy groups for the Boltzmann replica, and four for the other replicas. In both cases there is a group containing all the non-dimerized atoms, 
for the Boltzmann replica there is a group for the virtual sites of the dimers and one for the beads, regardless of their color. For the other replicas there are two distinct groups for 
each bead color and a group for the virtual sites. The interaction tables between (and within) these groups are automatically generated by the Dimerizer script for the usual 6-12 Lennard-Jones potential and $1/r$ Coulomb interaction. 

To help getting started with the Dimer Metadynamics, 
this package offers a code documentation and a directory containing the examples that are shown in the next section.

A slight modification to the forces has been necessary when using Particle-Mesh Ewald (PME) to treat the Coulomb interaction. Since Gromacs, at least 
up to the last version at the time of writing (5.1.5) does not allow a separate treatment of PME for each energy group, 
the charges and their relative interactions are always computed on the virtual sites, regardless of the replica index. This is correct for the classical replica, and in fact it is equivalent to the PME energy in a 
conventional simulation; since the other replicas are unphysical by definition, this choice is legitimate and has the advantage to converge to the classical energy as $\sigma \rightarrow 0$.

\section*{\sffamily \Large RESULTS}
\begin{figure}
\includegraphics[width=0.7\columnwidth,keepaspectratio=true]{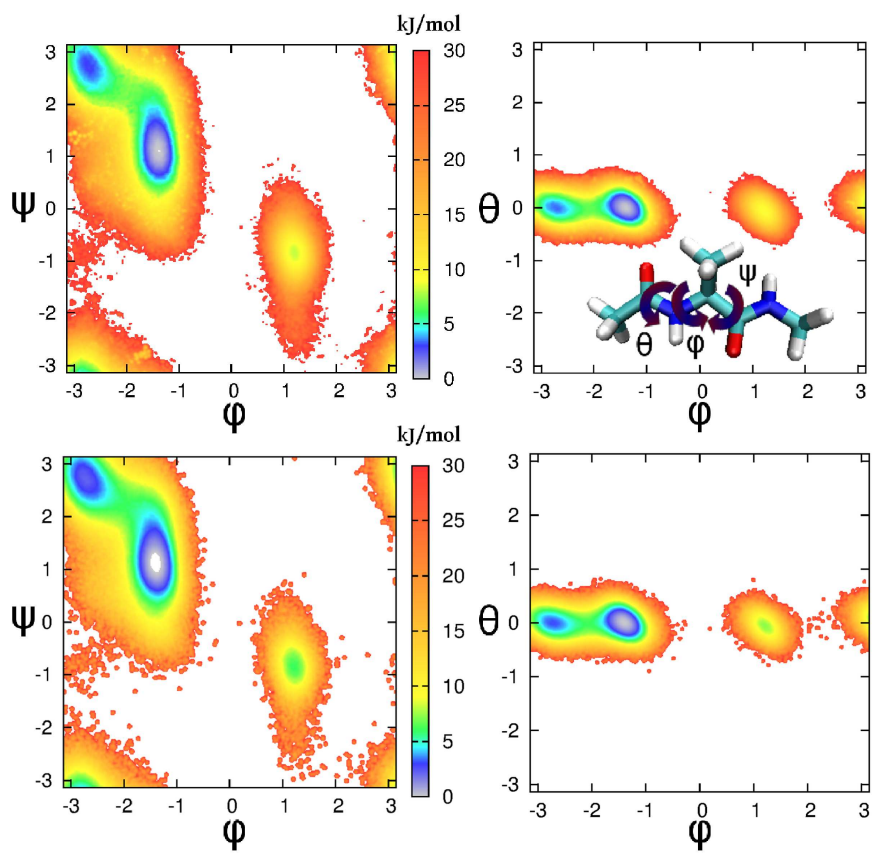}
\caption{\label{res_alav} (Color online) Free energy surface of the dihedral angles $\phi$,$\psi$ (right) and $\theta$,$\phi$ 
of Alanine Dipeptide in vacuum as defined in the inset. 
Upper panel are results obtained with Dimer Metadynamics on Gromacs, lower panel are the same results obtained with Parallel Tempering.
}
\end{figure}

We demonstrate the Gromacs implementation of Dimer Metadynamics by considering once again Alanine Dipeptide at $T=300$ K 
with and without water solvent. We have used four replicas in which only the backbone atoms were dimerized. 
For Alanine Dipeptide in vacuum the dimer interaction parameters were $\sigma_i = $ 0.002, 0.007 and 0.9 nm so that 
the acceptation probability of the swaps between replicas is about 0.2. The 
simulation length was 400 ns with a timestep of 1 fs. Well-tempered Metadynamics has been used with 
bias factor $\gamma = 10$, the Gaussians were deposited every 0.5 ps and had initial height 
of $h_0=0.5$ kJ/mol and widths $\sigma_G = $15, 15, 8, 0.0008 nm respectively for each replica. In Figure \ref{res_alav}(
upper panel), the free energy surface of the distribution of the dihedral angles $\theta,\phi,\psi$ is shown together with the definition of the angles. 
The results are compared with a Parallel Tempering simulation consisting of 3 replicas with 
temperatures $T_i=$300, 387.3 and 500 K that ran for 400 ns with a timestep of 2 fs. The agreement between the two different methods is 
very good and also the sampled regions are very similar for the whole free energy range (not shown in the picture). 
The degree of consistency of the comparison between Dimer and PT-WTE results can be quantified by computing the 
Battacharyya distance between the distributions and, for the $\phi,\psi$ distribution this value is 0.008.

We then moved on to test the method in the presence of a solvent. As expected, adding the solvent to Alanine Dipeptide did not require 
a larger number of replicas since the energy coupling of the protein with the solvent does not enter in the acceptation probability of the swaps. 
The replicas had interaction parameters $\sigma_i= $0.002, 0.005 and 1.0 nm and the simulation length was 80 ns with a timestep of 1 fs. 
The Metadynamics Gaussians were deposited every 0.5 ps, with initial height $h_0=0.5$ kJ/mol and $\sigma_G = 8, 5, 5, 0.0008$ nm. The bias factor was $\gamma = 10$.
Again, the results have been checked with a Parallel Tempering in the Well Tempered Ensamble\cite{wte} (PT-WTE) simulation consisting of 4 replicas at temperatures $T_i=$300.0, 
 349.9, 416.5 and 504.6 K, with Metadynamics 
bias factor $\gamma= 20$ and Gaussians of initial height $h_0=$2.5 kJ/mol and width $\sigma_G=200$ kJ/mol deposited every 1 ps. The simulation was 370 ns long, with a 
timestep $dt = 2$ fs. 

\begin{figure}
\includegraphics[width=0.7\columnwidth,keepaspectratio=true]{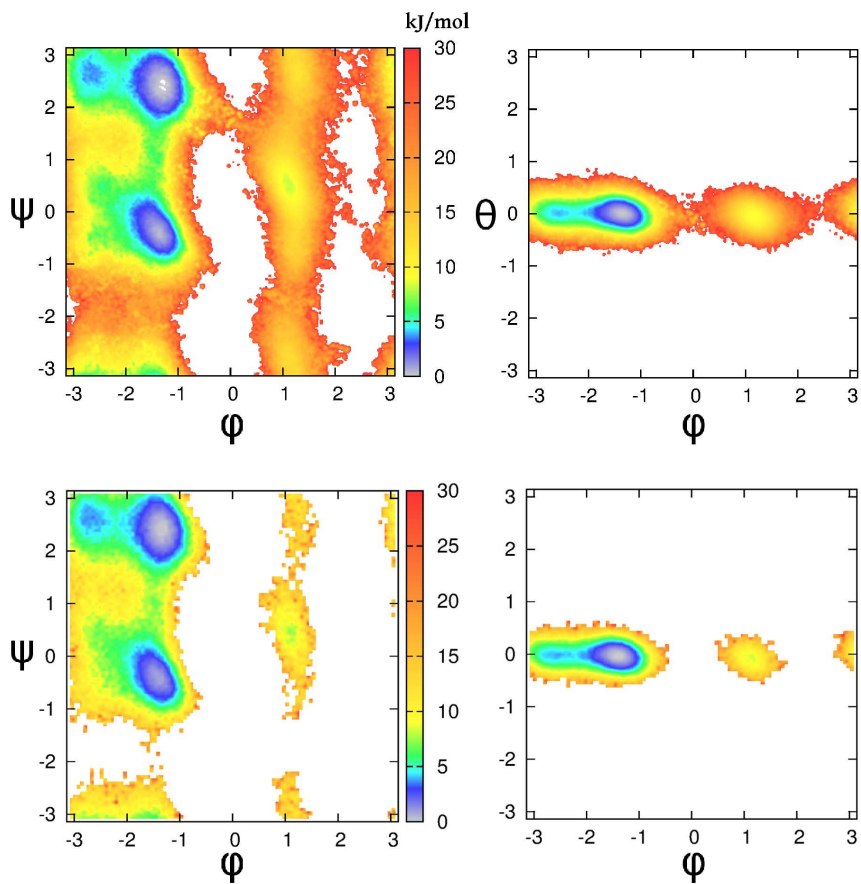}
\caption{\label{res_alaw} (Color online) Free energy surface of the dihedral angles $\phi$,$\psi$ (right) and $\theta$,$\phi$ of Alanine Dipeptide in water 
as defined in the inset. 
Upper panel are results obtained with Dimer Metadynamics on Gromacs, lower panel are the same results obtained with Parallel Tempering 
in the Well-Tempered Ensemble.
}
\end{figure}

The results are shown in figure \ref{res_alaw} for the dihedral angles $\phi$,$\psi$ and $\theta$, all the minima are represented consistently, 
the distance between the two distributions $\phi,\psi$ is in fact as small as 0.01. We note however that for PT-WTE (lower panel) the distributions are rougher even though the simulation was much longer. This is possibly an effect of the reweighting that in Dimer Metadynamics is less important since the 
Metadynamics bias on the classical replica acts on degrees of freedom that are decoupled from the ones that determine the results.

As a last example, to demonstrate the applicability of the method on larger systems we have considered the 12-residue Alanine Polypeptide 
in water, depicted in Figure \ref{res_alaw}. Only the backbone has been dimerized and it required 10 replicas of $\sigma_i=$0.0012, 0.0018, 0.0025,
 0.006, 0.01, 0.03, 0.1, 0.26 and 1.2 nm to obtain convergence in the analyzed properties. 
The simulation was 70 ns long, with a timestep of 0.5 fs. Metadynamics was used with a bias factor of $\gamma = 10$; the Gaussians were 
deposited every 250 ps and had initial height $h_0=0.5$ kJ/mol and widths $\sigma_G=$20, 20, 20, 10, 6, 3, 0.7, 0.08, 0.02 and 0.002 nm in order to be of size compatible of 
the fluctuations in dimer energy for each replica. 

The reference simulation with PT-WTE was 270 ns long, with a timestep of 1 fs and temperatures $T_i=$300, 320.3, 343.2, 368.7,
 397.1, 429, 464.9 and 505.5 K. Metadynamics was used to enhance the acceptation probability of the swaps between different temperature, the bias factor was $\gamma = 40$ and the Gaussians 
were deposited every 500 ps, with $h_0=2.5 $ kJ/mol and $\sigma_G=200 $kJ/mol. The comparison of the FES of the dihedral angles $\phi_6$ and $\psi_3$ is shown in 
figure \ref{res_alaw} along with the definition of the dihedral angles. Also in this case, the results are in good agreement, the distance between distributions for $\psi,\phi$ is 0.03. 
Dimer Metadynamics samples the same minima as in PT-WTE, like in the previous case not only the distribution is smoother, but it also required a simulation length shorter compared to PT-WTE. 
Moreover, with Dimer Metadynamics the regions between minima seem to be better sampled than in the PT case and can possibly provide information about the properties of the free energy barriers.
The number of replicas required for the Dimer simulation is slightly higher, however we note that this number could be reduced by increasing the bias factor to a value closer to 
what has been used for the PT-WTE simulation.

\begin{figure}
\includegraphics[width=0.7\columnwidth,keepaspectratio=true]{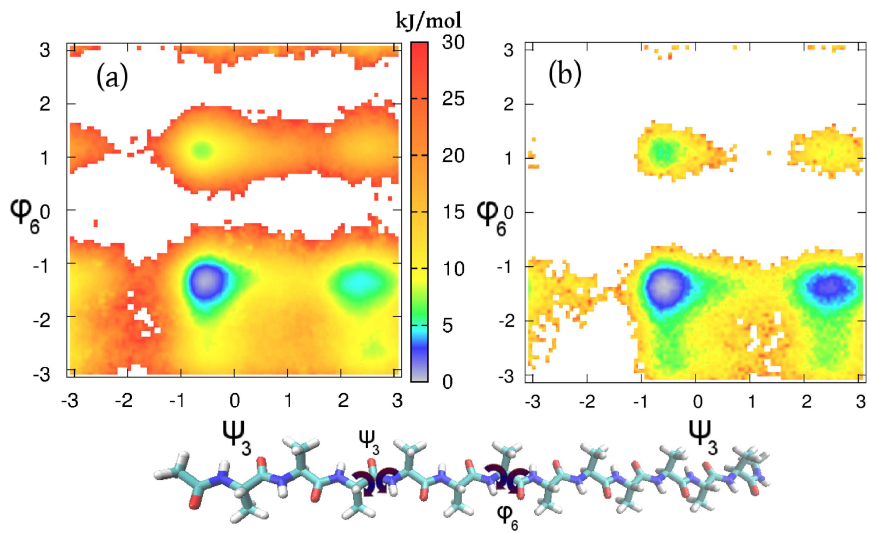}
\caption{\label{res_ala12} (Color online) Free energy surface of the dihedral angles $\phi_6$,$\psi_3$ of 12-residue Alanine Polypeptide in water 
as defined in the inset. (a) Results obtained with Dimer Metadynamics on Gromacs. (b) The same results obtained with Parallel Tempering 
in the Well-Tempered Ensemble.
}
\end{figure}

\section*{\sffamily \Large CONCLUSIONS}
In this work we have implemented Dimer Metadynamics\cite{gpi} in GROMACS. The implementation was obtained with a rather intricate input format that 
allowed us to avoid modifying the source code of the program, relieving thus the user from the often cumbersome task of compiling a custom version of the code. 
This required also the creation of modified forcefields and user non-bonded interaction tables, as well as ``dimerized'' topologies. 
All these intricacies have been taken care of with a easy to use Python script that has been made publicly available on Github (see acknowledgements section). 
We have demonstrated not only the Gromacs implementation but also the ability of Dimer Metadynamics to work even in the presence of an explicit solvent. 


\subsection*{\sffamily \large ACKNOWLEDGMENTS}
We acknowledge the Swiss National Science Foundation grant 200021\_169429/1 for funding. 
We would also like to thank Dr. Ferruccio Palazzesi for his assistence with PT simulations.

The Dimerizer package and relative documentation can be found on Github: 

\hspace{2cm}https://github.com/marckn/dimerizer

The Dimer CV is available in the official Plumed repository: 

\hspace{2cm}https://github.com/plumed/plumed2

\clearpage




\end{document}